\documentstyle[prc,preprint,aps,psfig]{revtex}

\tightenlines

\begin{document}
\title{Correlation energy of the pairing Hamiltonian}
\author{K. Hagino and G.F. Bertsch}
\address{Institute for Nuclear Theory, Department of Physics, \\
University of Washington, Seattle, WA 98195, USA}
\maketitle

\bigskip

\begin{abstract}

We study the correlation energy associated with the pair fluctuations 
in BCS theory. 
We use a schematic two-level pairing model and discuss 
the behavior of the correlation energy across shell closures, including 
the even-odd differences. 
It is shown 
that the random phase approximation (RPA) to the ground state 
energy reproduces the exact solutions quite well  
both in the normal fluid and in the superfluid phases. 
In marked contrast, other methods to improve the BCS energies, such as 
Lipkin-Nogami method, only work when the strength 
of a pairing interaction is large. 
We thus conclude that the RPA approach 
is better for a systematic theory of nuclear binding energies. 

\end{abstract}   


\section{Introduction}

An important goal of nuclear theory is to predict
nuclear binding energies.  
The mean-field approximation certainly provides a good starting
point, but the correlation energy associated with nearly degenerate 
configurations can be of the order of several MeV. 
Since one would 
like an order of magnitude better accuracy for global modeling,
the correlations must be treated with some care.
The correlations associated with broken mean-field symmetries, namely
center-of-mass localization, deformations, and pairing, are especially
important and need to be singled out for special attention.
The study of such correlation effects  
has attracted much interest, see Refs. \cite{BRRM99,VHB99,RER00} 
for recent citations. 
Popular computational methods
\footnote{One may also ask whether simplified exact solutions
may be of use for realistic applications.  
The Richardson solution\cite{RS64,R77} 
for a pairing Hamiltonian is useful in this respect, 
but it requires
a state-independent pairing interaction, which may not be realistic
enough.  That solution has recently been generalized to separable
pairing interactions\cite{PD98}, but it is not yet clear whether the
computational simplicity is preserved.} 
include the generator coordinate method (GCM)\cite{RS80}, 
variation-with-projection
methods (e.g., VAP) \cite{BFV99}, and the RPA \cite{R68}.  

In previous work\cite{HB00}, we have advocated the RPA approach for 
deformations and for center-of-mass localization. 
The correlation energy is calculated from the RPA
formula \cite{R68}
\begin{equation}
E_{corr}= \frac{1}{2}\left(\sum_i\hbar\omega_i - Tr(A)\right), 
\label{corr}
\end{equation}
where $\omega_i$ is the frequency of the RPA phonon 
and $A$ is the $A$ matrix in the RPA equations.  This method requires
solving the RPA equations for each nucleus, which is computationally
easy if the interaction is assumed to be separable\cite{DAN99,SDB99,JBH99}.  
On other other hand, methods such as VAP and GCM are rather complicated to
apply.  One should also remember that a global theory requires 
a method that is 
not only systematic but also preserves the relative computational simplicity
of the mean field approximation.  In this respect, the Lipkin-Nogami
method is a popular candidate for a computationally easy method to
go beyond the mean field pairing theory\cite{L60,N64,PNL73} .

In this paper, we continue our study of the accuracy of the
RPA correlation formula 
(\ref{corr}), going on to pairing correlations.  There is a long history
of the RPA treatment of pairing correlations.  An early study by
Bang and Krumlinde \cite{BK70} showed that the RPA formula reproduces
the exact correlation energy rather well in a schematic model.  Kyotoku
{\it et al.} \cite{KLC96} compared the leading $1/N$ behavior of different
methods including the Lipkin-Nogami method.  They found that
only RPA gave the exact coefficient in the condensed phase.
The RPA method has in fact been used in realistic models of deformed
nuclei\cite{S89}.  
The RPA correlation in the normal phase was studied in Ref. \cite{DRS98} 
using the self-consistent version of RPA. 
But perhaps surprisingly in view of the large
literature, we have not found any studies that specifically compare
the RPA with the computationally attractive alternative methods,
testing the behavior across shell closures and for odd $N$ systems.  
It is important that 
the method should not introduce spurious discontinuities when the
mean field solution changes character; otherwise separation energies
could not be reliably calculated \cite{Nazarewicz}. 

To investigate the applicability of Eq. (\ref{corr}) and study the effects 
of symmetry breaking restoration, in this paper 
we employ a well-known schematic two-level 
pairing model\cite{PNL73,DRS98,HF61,CP90,ZSF92,CR98,PTK98,SD99}, 
and compare its exact solutions to the
approximations that are computationally attractive.  The details of
the RPA correlation energy are given in the next section.  
They include both the 
quasi-particle RPA extension of the BCS theory 
and the RPA for the pairing
Hamiltonian when the strength is too weak for the mean field approximation
to support the BCS solution (the ``pairing-vibration" regime).  
The specific application
of the various methods to the two-level model is given in the following
section III.

\section{Pairing Hamiltonian and the RPA}

The pairing Hamiltonian and the BCS solution are well-known and we
just summarize the equations to confirm the standard notation.  
In this paper we consider a Hamiltonian with an arbitrary single-particle
term but a pairing 
interaction whose strength is state-independent,
\begin{equation}
H=\sum_j\epsilon_j \hat{N}_j-G\sum_{j,j'}A_j^{\dagger}A_{j'},
\label{Hpair}
\end{equation}
where $\epsilon_j$ is a single-particle energy. The number operator
$\hat{N}_j$ for the shell $j$ is given by
\begin{equation}
\hat{N}_j=\sum_{m > 0}\left(a_{jm}^{\dagger}a_{jm} + a^{\dagger}_{j\bar{m}}
a_{j\bar{m}}\right).
\end{equation}
Here, $a_{j\bar{m}}$ is the annihilation operator for the time-reversal 
state and is given by $a_{j\bar{m}}=(-)^{j-m}a_{j-m}$. 
The pairing operator $A_j^{\dagger}$ in the Hamiltonian (\ref{Hpair}) 
is given by 
\begin{equation}
A_j^{\dagger}
=\sum_{m > 0}a^{\dagger}_{jm} a^{\dagger}_{j\bar{m}},
\end{equation}
and $A_j$ is the Hermitian conjugate of  $A_j^{\dagger}$. 
The exact solutions of the Hamiltonian were obtained a long time ago by 
Richardson and Sherman\cite{RS64}.
In Appendix A, we solve the Richardson equation 
for a simple, but non-trivial case where there are two pairs in a single 
$j$-shell. 

The BCS theory is the mean field solution to Eq. (\ref{Hpair}). We remind 
the reader that it is derived variationally from a ground state wave
function of the form
\begin{equation}
|BCS\rangle
=\prod_{j,m>0}(u_j+v_j a^{\dagger}_{jm} a^{\dagger}_{j\bar{m}})|\rangle,
\end{equation}
where $v_j$  is the usual pair occupation amplitude and 
$u_j$ satisfies $u_j^2+v_j^2=1$. 
This wave function is appropriate for even-$N$ systems, $N$ being the number 
of particles in a system. The 
extension to odd-$N$ systems is given at the end of 
this subsection. 
Important quantities in the variational
solution are the chemical potential $\lambda$ and the pairing gap 
$\Delta=G\sum_j\Omega_ju_jv_j$, $\Omega_j=(2j+1)/2$ being the pair degeneracy 
of the $j$-shell.  When the pairing strength $G$ is small, the variational
equations only have the trivial solution, $v_j,u_j=0$ or 1 and $\Delta=0$.
The ground state wave function is thus given by 
$|HF\rangle = \prod_{j,m}a_{jm}^{\dagger}|\rangle$ and the ground state 
energy is obtained as
\begin{equation}
E_{HF}=\langle HF|H|HF\rangle=\sum_{j:occupied}\Omega_j(2\epsilon_j-G).
\end{equation}
When the strength of the pairing interaction $G$ is larger than some critical 
value $G_{crit}$, the variational equations have a non-trivial 
solution 
$\Delta \neq 0$ and the ground state energy is given by 
\begin{equation}
E_{BCS}=\langle BCS|H|BCS\rangle=2\sum_{j}\Omega_j\epsilon_j v_j^2
-\frac{\Delta^2}{G}-G\sum_j\Omega_jv_j^4.
\label{ebcs}
\end{equation}

For odd $N$ systems, one of the particles in a system does not form a pair 
and blocks a level. When the $k$-th level is blocked, the HF energy, the 
BCS energy, and the pairing gap are modified to 
\begin{eqnarray}
E_{HF}&=&\sum_{j:occupied}\widetilde{\Omega}_j
(2\epsilon_j-G) + \epsilon_k, \\
E_{BCS}&=&2\sum_{j}\widetilde{\Omega}_j\epsilon_j v_j^2
-\frac{\Delta^2}{G}-G\sum_j\widetilde{\Omega}_jv_j^4+ \epsilon_k, \\
\Delta&=&G\sum_j\widetilde{\Omega}_ju_jv_j,
\end{eqnarray}
respectively. Here, $\widetilde{\Omega}_j = \Omega_j$ for $j\neq k$, and 
$\widetilde{\Omega}_k = \Omega_k -\delta_{N,odd}$. 
The chemical potential $\lambda$ is determined so that 
\begin{equation}
\sum_j\widetilde{\Omega}_jv_j^2+1=N,
\end{equation}
is satisfied. 
This formalism is referred to as 
the blocked-BCS theory. 

\subsection{Random phase approximation}


We next introduce the random phase approximation to compute the correlation 
energy. 
Let us first consider the RPA in the superfluid phase (QRPA). 
We refer to Refs. \cite{RS80,HF61,J70} for the formulation. 
The result is the well-known RPA matrix equation 
\begin{equation}
\left(\begin{array}{cc}
A & B \\
-B & -A 
\end{array}\right)
\left(\begin{array}{c}
X \\ Y
\end{array}\right)
= \hbar\omega \left(\begin{array}{c}
X  \\ Y
\end{array}\right),
\label{QRPA}
\end{equation}
where the matrices $A$ and $B$ are given by
\begin{eqnarray}
A_{ij}&=&2E_i\delta_{i,j}-G\sqrt{\widetilde{\Omega}_i}
\sqrt{\widetilde{\Omega}_j}(u_i^2u_j^2
+v_i^2v_j^2), \label{QRPA-A} \\
B_{ij}&=&G\sqrt{\widetilde{\Omega}_i}\sqrt{\widetilde{\Omega}_j}(u_i^2v_j^2
+v_i^2u_j^2),
\label{QRPA-B}
\end{eqnarray}
respectively. 
Here, $\widetilde{\Omega}_j$ is defined in the previous subsection, and 
$E_j = \sqrt{(\epsilon_j-\lambda-Gv_j^2)^2+\Delta^2}$ is the quasi-particle 
energy. 
The RPA excitation operator $Q^{\dagger}$ is given in terms of 
quasi-particle operators  
\begin{eqnarray}
\alpha_{jm}^{\dagger}&=&u_ja^{\dagger}_{jm}-v_ja_{j\bar{m}}, \\
\alpha_{j\bar{m}}^{\dagger}&=&u_ja^{\dagger}_{j\bar{m}}+v_ja_{jm},
\end{eqnarray}
as 
\begin{equation}
Q^{\dagger}=\sum_j
\left(X_j \sum_m\alpha_{jm}^{\dagger}\alpha_{j\bar{m}}^{\dagger}
-Y_j\sum_m\alpha_{j\bar{m}}\alpha_{jm}\right)/\sqrt{\widetilde{\Omega}_j}.
\end{equation}
The QRPA correlation energy is given 
by Eq. (\ref{corr}) with the $A$ matrix given by Eq. (\ref{QRPA-A}). 
Note that the second term on the right hand side of 
Eq. (\ref{corr}) is not $\sum_i 2E_i$ but $Tr(A)$. This point was overlooked 
in the integral approaches to the correlation energy in Refs. 
\cite{DAN99,SDB99}. 

We next consider the RPA in a normal-fluid phase (pp-RPA), 
which describes pairing vibrations. 
The equation for the pp-RPA can be obtained from the QRPA equation by 
setting $\Delta=0$ and $v_h=u_p=1$, where $p$ and $h$ denote 
particle and hole states, respectively. 
As for the chemical potential $\lambda$, there is no definite value for 
it in the normal fluid phase since it can be anywhere between the highest 
occupied and the lowest unoccupied levels. 
Notice, however, that the pairing vibration describes 
the ground state in the $N\pm2$ systems and the role of the chemical 
potential is therefore just to shift the energies by an amount 
$\pm 2\lambda$. 
After removing these trivial energy shifts, one obtains
\begin{eqnarray}
A_{pp'}&=&2\epsilon_p\delta_{p,p'}-G\sqrt{\widetilde{\Omega}_p}
\sqrt{\widetilde{\Omega}_{p'}}, \\
A_{hh'}&=&-2(\epsilon_h-G)\delta_{h,h'}-G\sqrt{\widetilde{\Omega}_h}
\sqrt{\widetilde{\Omega}_{h'}}, \\
B_{ph}&=&G\sqrt{\widetilde{\Omega}_p}\sqrt{\widetilde{\Omega}_h}, \\
A_{ph}&=&B_{pp'}=B_{hh'}=0.
\end{eqnarray}
This 
$2(N_p+N_h)\times 2(N_p+N_h)$ 
dimensional matrix equation, where $N_p$ is the number of 
unoccupied shells and 
$N_h$ is that of occupied shells, 
can be decoupled into two $(N_p+N_h)\times (N_p+N_h)$ matrix equations as 
\begin{equation}
\left(\begin{array}{cc}
A & -B \\
B & -C 
\end{array}\right)
\left(\begin{array}{c}
X^a \\ Y^a
\end{array}\right)
= \hbar\omega_a \left(\begin{array}{c}
X^a  \\ Y^a
\end{array}\right),
\label{ppRPAa}
\end{equation}
and 
\begin{equation}
\left(\begin{array}{cc}
C & -B \\
B & -A 
\end{array}\right)
\left(\begin{array}{c}
X^r \\ Y^r
\end{array}\right)
= \hbar\omega_r \left(\begin{array}{c}
X^r  \\ Y^r
\end{array}\right). 
\label{ppRPAr}
\end{equation}
Here, the matrix $C$ is defined as $C_{hh'}=A_{hh'}, C_{ph}=0$, and 
RPA amplitudes are given by $X^a_p=X_p, Y^a_h=-Y_h, X^r_h=-X_h$, and 
$Y^r_p=Y_p$. The first equation (\ref{ppRPAa}) describes 
the ground state of the $N+2$ system and is referred to as the addition mode, 
while the second equation (\ref{ppRPAr}) describes the $N-2$ system 
and is called the removal mode. 
Noticing that Eq. (\ref{corr}) is 
a half of a sum of the difference between the RPA and the Tamm-Dancoff 
approximation (TDA) frequencies for each mode of excitation \cite{RS80}, 
we find the correlation energy for the addition mode to be 
\begin{equation}
E^a_{corr}= \frac{1}{2}\left(\sum_i\hbar\omega_{ai} - Tr(A)\right), 
\end{equation}
while that for the removal mode is 
\begin{equation}
E^r_{corr}= \frac{1}{2}\left(\sum_i\hbar\omega_{ri} - Tr(C)\right). 
\end{equation}
The total correlation energy is the sum of these, 
$E_{corr}=E^a_{corr}+E^r_{corr}$. 

Typically, one finds that an RPA mode goes to zero frequency at the 
mean-field phase transition. This is not the case for Eqs. (\ref{ppRPAa}) 
and (\ref{ppRPAr}), thus giving a discontinuity in the 
frequencies at the transition. However, as we 
have mentioned above, this is an artifact of an awkward choice of 
the chemical potential $\lambda$, and 
it will be the case 
that the sum of an addition and removal mode goes to zero. 

\subsection{Lipkin-Nogami method}

An alternative way to restore the broken gauge symmetry of the BCS 
approximation is to carry out the number projection of the 
BCS wave function. The Lipkin-Nogami method provides an approximate 
way for number projection. It was first invented by Lipkin \cite{L60}, 
and was developed by Nogami and his collaborators \cite{N64,PNL73}. 
Because of its relative simplicity, it has been widely applied 
\cite{ZSF92,RNBM96,MN92,BR96,QRMM90}. 

In the Lipkin-Nogami method, the expectation value 
$\langle H-\lambda \hat{N} -\lambda_2\hat{N}^2\rangle_{BCS}$ 
is varied. 
The resultant equations to be solved are given by \cite{PNL73}
\begin{eqnarray}
&&\frac{2}{G}=\sum_j\frac{\widetilde{\Omega}_j}
{\sqrt{\tilde{\epsilon}_j+\Delta^2}}, \\
&&2\sum_j\widetilde{\Omega}_j v_j^2 + \delta_{N,odd}= N, \\
&&\lambda_2=\frac{G}{4}\left\{
\frac{\left(\sum_i\widetilde{\Omega}_iu_i^3v_i\right)
\left(\sum_i\widetilde{\Omega}_iu_iv_i^3\right)
-\sum_i\widetilde{\Omega}_iu_i^4v_i^4}
{\left(\sum_i\widetilde{\Omega}_iu_i^2v_i^2\right)^2
-\sum_i\widetilde{\Omega}_iu_i^4v_i^4}\right\},
\end{eqnarray}
where $\tilde{\epsilon}_i, v_i$, and $u_i$ are defined as
\begin{eqnarray}
\tilde{\epsilon}_i &=& \epsilon_i+(4\lambda_2-G)v_i^2-\lambda, \\
v_i^2&=&\frac{1}{2}\left(1-\frac{\tilde{\epsilon}_i}
{\sqrt{\tilde{\epsilon}_i^2+\Delta^2}}\right), \\
u_i^2&=&1-v_i^2,
\end{eqnarray}
respectively. 
We use the blocked-Lipkin-Nogami prescription for odd $N$ systems. 
One of the characteristic features of the Lipkin-Nogami method is that 
the pairing gap $\Delta$ has a finite value even in the weak $G$ limit where 
$\Delta$ is zero in the BCS approximation. 
The ground state energy in the Lipkin-Nogami method is given by 
\begin{equation}
E_{LN}=2\sum_{j}\widetilde{\Omega}_j\epsilon_j v_j^2
-\frac{\Delta^2}{G}-G\sum_j\widetilde{\Omega}_jv_j^4
-4\lambda_2\sum_j\widetilde{\Omega}_ju_j^2v_j^2 + \epsilon_k\delta_{N,odd}.
\label{eln}
\end{equation}

\section{Two-level pairing model}

We now apply the above equations to a schematic two-level model. 
This model was first introduced in Ref.\cite{HF61}, and 
has been used in the literature to test several approximations 
\cite{PNL73,KLC96,DRS98,CP90,ZSF92,CR98,PTK98,SD99}. 
We label the lower and the higher levels 1 and 2, taking 
$\epsilon_1=-\epsilon/2, \epsilon_2=\epsilon/2$. 
The Hamiltonian can be numerically diagonalised using the quasi-spin 
formalism \cite{RS80}. The basis states are denoted by 
$|S_1S_{10}; S_2S_{20}\rangle$, where $S_i$ and $S_{i0}$ are 
defined by $S_i=(\Omega_i-\nu_i)/2$ and 
$S_{i0}=(N_i-\Omega_i)/2$, respectively. The latter 
takes a value of $-S_i, -S_i+1, \cdots, 
S_i$. $\nu_i$ is the seniority quantum number. For the ground state of 
even-$N$ systems, the total seniority $\nu_1+\nu_2$ is 0, while it is 1 for 
odd-$N$ systems. 
The matrix elements of the Hamiltonian read 
\begin{eqnarray}
&&\langle S_1'S_{10}'; S_2'S_{20}'|H|S_1S_{10}; S_2S_{20}\rangle
=\delta_{S_1,S_1'}\delta_{S_2,S_2'}\left\{
\epsilon\left(S_{20}-S_{10}-\frac{\Omega_1-\Omega_2}{2}\right)
\delta_{S_{10},S_{10}'}\delta_{S_{20},S_{20}'}\right. \nonumber \\
&& \quad\quad 
-G\left(S_1(S_1+1)-S_{10}(S_{10}+1)+S_2(S_2+1)-S_{20}(S_{20}+1)\right)
\delta_{S_{10},S_{10}'}\delta_{S_{20},S_{20}'} \nonumber \\
&& \quad\quad 
-G\sqrt{(S_1(S_1+1)-S_{10}(S_{10}-1)}\sqrt{S_2(S_2+1)-S_{20}(S_{20}+1)}
\delta_{S_{10},S_{10}'+1}\delta_{S_{20},S_{20}'-1} \nonumber \\
&& \quad\quad \left.
-G\sqrt{(S_1(S_1+1)-S_{10}(S_{10}+1)}\sqrt{S_2(S_2+1)-S_{20}(S_{20}-1)}
\delta_{S_{10},S_{10}'-1}\delta_{S_{20},S_{20}'+1} \right\}.
\end{eqnarray}
 
We first consider a symmetric two-level problem i.e., 
$\Omega_1=\Omega_2=\Omega$, and assume the number of 
fermionic particle is $N=2\Omega$. Equations for several approximations 
can be solved analytically for such a system. 
The gap equation in the BCS theory leads to a pairing gap of 
\begin{equation}
\Delta=\sqrt{G^2\Omega^2-\frac{\tilde{\epsilon}^2}{4}},
\label{gap2}
\end{equation}
together with 
\begin{eqnarray}
v_1^2&=&u_2^2=\frac{1}{2}\left(1+\frac{\tilde{\epsilon}}{2G\Omega}\right)
\equiv v^2, 
\label{v2} \\
u_1^2&=&v_2^2=\frac{1}{2}\left(1-\frac{\tilde{\epsilon}}{2G\Omega}\right)
\equiv u^2. \\
\lambda&=&\frac{\tilde{\epsilon}-\epsilon}{2}-Gv^2, 
\label{chem}
\end{eqnarray}
where $\tilde{\epsilon}$ is defined as 
$\tilde{\epsilon}=2\Omega\epsilon/(2\Omega-1)$. 

From Eq. (\ref{gap2}), one can find the critical strength of the 
phase transition to be $G_{crit}=\epsilon/(2\Omega-1)$. 
For $G$ larger than $G_{crit}$, the system is in the superfluid phase, and 
the matrices $A$ and $B$ for the QRPA equation 
are given by (see Eqs. (\ref{QRPA-A}) and (\ref{QRPA-B}))
\begin{eqnarray}
A_{11}&=&A_{22}=
G\Omega
-\frac{\Delta^2}{2G\Omega^2}
+\frac{\Delta^2}{2G\Omega}, \\
A_{12}&=&A_{21}=-\frac{\Delta^2}{2G\Omega}, \\
B_{11}&=&B_{22}=
-\frac{\Delta^2}{2G\Omega^2}
+\frac{\Delta^2}{2G\Omega}, \\
B_{12}&=&B_{21}=
G\Omega-\frac{\Delta^2}{2G\Omega}, 
\end{eqnarray}
The solutions of the QRPA equation are $\hbar\omega=0$ and 
$\sqrt{4\Delta^2-2\Delta^2/\Omega}$, from 
which the correlation energy is computed as
\begin{equation}
E_{corr}=
\sqrt{\Delta^2-\Delta^2/2\Omega}
-G\Omega
+\frac{\Delta^2}{2G\Omega^2}
-\frac{\Delta^2}{2G\Omega}. 
\end{equation}

When the strength of the pairing interaction $G$ is smaller than 
$G_{crit}$, the system is in the normal fluid phase. 
The mean field energy is then evaluated as 
$E_{HF}=-\epsilon\Omega-G\Omega$.
For the present two-level model, the $A, B$, and $C$ matrices for the 
pp-RPA are just numbers and are given by
$A=\epsilon-G\Omega$, $B=G\Omega$, and $C=\epsilon-G\Omega+2G$, respectively. 
The frequencies for the addition and the removal modes are then 
found to be 
\begin{eqnarray}
\hbar\omega_a&=&-G+\sqrt{\epsilon+G}\sqrt{\epsilon+G-2G\Omega}, \\
\hbar\omega_r&=&G+\sqrt{\epsilon+G}\sqrt{\epsilon+G-2G\Omega}, 
\end{eqnarray}
respectively. 
The total correlation energy is thus obtained as 
\begin{equation}
E_{corr}=\sqrt{\epsilon+G}\sqrt{\epsilon+G-2G\Omega}
-(\epsilon-G\Omega+G).
\end{equation}

Figure 1 shows the RPA frequencies for each modes of excitation 
as a function of $G$, for $\Omega$ = 8. 
As we noted in the previous section, 
we see that because of the chemical potential the pp-RPA frequencies 
do not match with the QRPA frequencies at the critical point of phase 
transition from a normal-fluid to a superfluid phases. 
Figure 2 compares the ground state energy obtained by several methods. 
The solid line is the exact solution obtained by numerically 
diagonalizing the Hamiltonian. The dashed line is the ground state energy 
in the mean field BCS approximation. 
It considerably deviates from the exact solution through 
the entire range of $G$ shown in the figure. The dot-dashed line 
takes into account the RPA correlation energy in addition to the mean field 
energy. It reproduces very well the exact solutions, except 
in the vicinity of the critical point of the phase transition. 
Around the critical point, one would need to 
compute the correlation energy with some care, using e.g., the self-consistent 
RPA discussed in Ref. \cite{DRS98} which removes the cusp behaviour of 
the RPA frequency around the critical strength. 

It is interesting to compare the present mean field plus RPA approach with 
the Lipkin-Nogami method. The equations for the Lipkin-Nogami method 
for the two-level model were solved in Ref. \cite{PNL73}. The results are 
given by
\begin{eqnarray}
\Delta^2&=&(G\Omega)^2(1-\tilde{\kappa}^2), \\
v_1^2&=&u_2^2=(1+\tilde{\kappa})/2, \\
u_1^2&=&v_2^2=(1-\tilde{\kappa})/2, \\
4\lambda_2-G&=&\frac{2G\Omega\tilde{\kappa}^2}
{(2\Omega-1)(1-\tilde{\kappa}^2)},
\end{eqnarray}
where $\tilde{\kappa}$ is the physical solution, which satisfies 
$0\leq \tilde{\kappa} \leq 1$, of an equation 
\begin{equation}
2(1-\Omega)\tilde{\kappa}^3+(2\Omega-1)\kappa\tilde{\kappa}^2
+(2\Omega-1)\tilde{\kappa}-(2\Omega-1)\kappa=0,
\end{equation}
and $\kappa$ is defined as $\epsilon/2G\Omega$. 
The ground state energy in the Lipkin-Nogami method given by Eq. (\ref{eln}) 
is denoted by the thin solid line in Fig. 2. Although it reproduces the 
exact results for large values of $G$, it deviates significantly from them 
for small values. 
This behaviour is consistent with 
the numerical observation in Ref. \cite{ZSF92} as well as 
the result of Ref.\cite{KLC96}
where it was found that the Lipkin-Nogami method is only correct in the 
limit of a strong pairing force. In marked contrast, 
the ground state energy in the BCS plus QRPA coincides with the exact solution 
at the leading order of an expansion in 1/$\Omega$ for any strength of 
the pairing interaction as long as it is larger than the critical 
value \cite{KLC96}. 

From Fig. 2, it is unclear whether the Lipkin-Nogami method or RPA are 
more accurate for purposes of a global theory of binding. 
So we now consider a more realistic situation, varying the particle number 
$N$ rather than the interaction strength $G$. We consider the paring energy in 
oxygen isotopes, taking the neutron 1p and 2s-1d shells as the 
lower and higher levels of the two-level model. 
The pair degeneracy $\Omega$ thus reads $\Omega_1=3$ and $\Omega_2=6$, and 
the number of particle in a system is given by $N=A-8-2$ for the $^A$O  
nucleus. We assume that the energy difference between the two levels 
$\epsilon$ is given by $\epsilon=41A^{-1/3}$ and the pairing strength 
$G=23/A$. 
The upper panel of Fig. 3 shows the ground state energy as a function of 
$A$. In order to match with the experimental data for the $^{16}$O nucleus, 
we have added a constant $-72.8$ MeV to the Hamiltonian for all the isotopes. 
The exact solutions are denoted by the filled circles. The deviation from 
the BCS approximation (the dashed line) is around 
2 MeV for even $A$ systems and it is around 1.2 MeV for odd $A$ systems. 
This value varies within about 0.5 MeV along the isotopes and shows relatively 
strong $A$ dependence. 
One can notice that the RPA approach (the dot-dashed line) 
reproduces quite well the exact solutions. 
On the contrary, the Lipkin-Nogami approach (the 
thin solid line) is much less satisfactory and shows 
a different $A$ dependence from the exact results. 
The pairing gaps $\Delta$ in the BCS approximation and in the Lipkin-Nogami 
method are shown separately in the lower panel of Fig. 3. 
For the Lipkin-Nogami 
method, we show $\Delta+\lambda_2$, which is to be compared with 
experimental data \cite{N64,PNL73}. The closed shell 
nucleus $^{16}$O and its neighbour nuclei $^{15,17}$O have a zero pairing 
gap in the BCS approximation, and the Lipkin-Nogami method does not 
work well for these nuclei, as can be casted in Fig. 2. 
On the other hand, the RPA approach reproduces the correct $A$ dependence 
of the binding energy. Evidently,  
the RPA formula provides a 
better method to compute correlation energies than the Lipkin-Nogami method, 
especially for shell closures. 

\section{Summary}

Returning to our initial motivation, we seek a computationally
tractable way to include pairing effects in a global model of 
nuclear binding energies, going beyond the BCS theory.  Two attractive
possibilities are the Lipkin-Nogami method and the RPA, in particular
if the pairing interaction has a separable form.  
In this paper, we used a solvable two-level pairing Hamiltonian to show that 
the RPA formula for the correlation energy reproduces well the exact solutions 
both in a normal-fluid and a superfluid phases. On the contrary, the 
Lipkin-Nogami method is considerably less accurate for weak pairing,
and therefore is not suitable in transition regions.
As a consequence, the Lipkin-Nogami method fails to reproduce the correct 
mass number dependence of the binding energy around a shell closure. 
The correlation energy 
associated with the number fluctuation for the neutron mode in O isotopes was 
shown to be order of 2 MeV and has a relatively strong mass dependence. 
Although the correlation energy is small compared with the absolute value 
of the ground state energy, this suggests that including the correlation 
energy in the RPA provides a promising way to develop a better microscopic 
systematic theory for nuclear binding energies. 

\section*{Acknowledgments}
The authors thank P.-G. Reinhard and W. Nazarewicz for useful discussions. 
This work was supported by the U.S. Department of Energy 
under Grant DOE-ER-40561. 

\begin{appendix}
\section{Richardson solution}

For a system where the number of pairs is $N_{pair}$, i.e., 
a 2$N_{pair}$-fermion system, the ground state energy of the Hamiltonian 
(\ref{Hpair}) is given by \cite{R77,PD98}
\begin{equation}
E_{gs}=\sum_{\lambda=1}^{N_{pair}}z_{\lambda},
\end{equation}
where $z_{\lambda}$ are $N_{pair}$ solutions of 
$N_{pair}$ coupled equations given by
\begin{equation}
\sum_j\frac{\Omega_j}{2\epsilon_j-z_{\lambda}}
-\mathop{{\;\,{\sum}'}}_{\lambda'} \frac{2}{z_{\lambda'}-z_{\lambda}}
=\frac{1}{G}. 
\end{equation}
The prime in the summation of $\lambda'$ means to take only those 
$\lambda'$ which are different from $\lambda$, and $\Omega_j = (2j+1)/2$ 
is the pair degeneracy of the $j$-shell. In general, $z_{\lambda}$ are 
complex. 

For a two pair system ($N_{pair}$=2) in a single $j$-shell, 
setting $\epsilon_j=0$ and $\Omega_j=\Omega$, the Richardson equation reads 
\begin{eqnarray}
-\frac{\Omega}{z_1}-\frac{2}{z_2-z_1}&=&\frac{1}{G} \\
-\frac{\Omega}{z_2}-\frac{2}{z_1-z_2}&=&\frac{1}{G}. 
\end{eqnarray}
The solutions of these equations are found to be 
\begin{equation}
z=-G(\Omega-1)\pm i\sqrt{G^2(\Omega-1)},
\end{equation}
from which the ground state energy reads $E_{gs}=-2G(\Omega-1)$. 
This result coincides with the solution obtained using the seniority scheme, 
\begin{equation}
E(N_{pair})=-G\left\{\frac{\Omega}{2}\left(\frac{\Omega}{2}+1\right)
-\frac{1}{4}(2N_{pair}-\Omega)^2+\frac{1}{2}(2N_{pair}-\Omega)\right\}.
\end{equation}

\end{appendix}

\newpage

\begin{figure}
  \begin{center}
    \leavevmode
    \parbox{0.9\textwidth}
           {\psfig{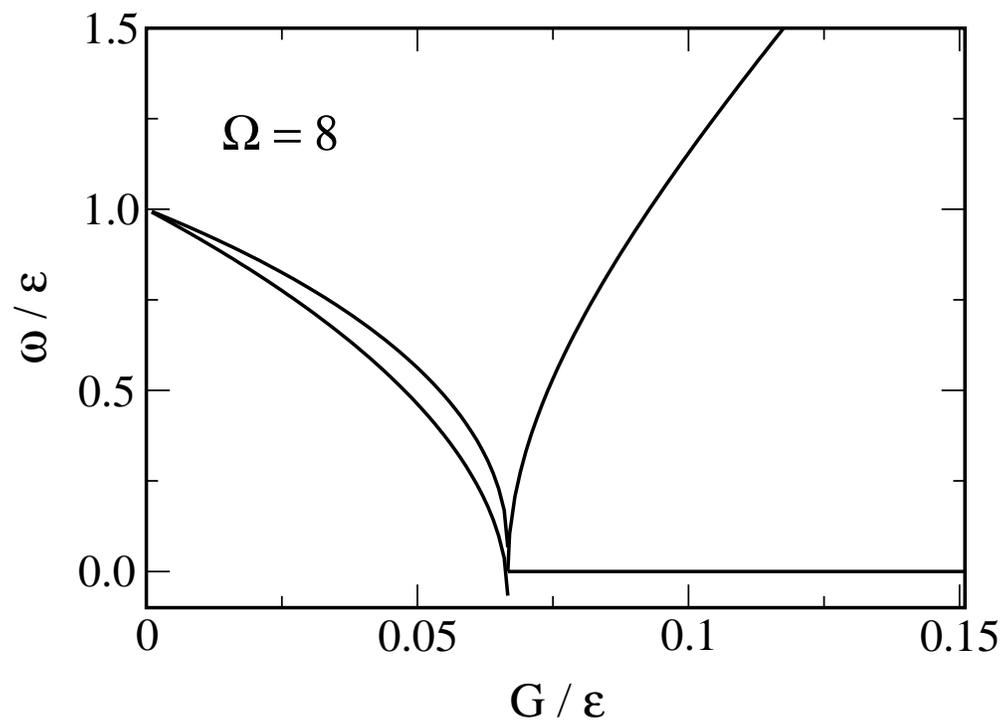}}
  \end{center}
\protect\caption{
RPA frequencies as a function of the strength of the pairing 
interaction $G$ for a two-level system. 
The pair degeneracy $\Omega$ 
is set to be 8. The strength of the pairing interaction $G$ 
is given in the unit of $\epsilon$. 
}
\end{figure}

\newpage

\begin{figure}
  \begin{center}
    \leavevmode
    \parbox{0.9\textwidth}
           {\psfig{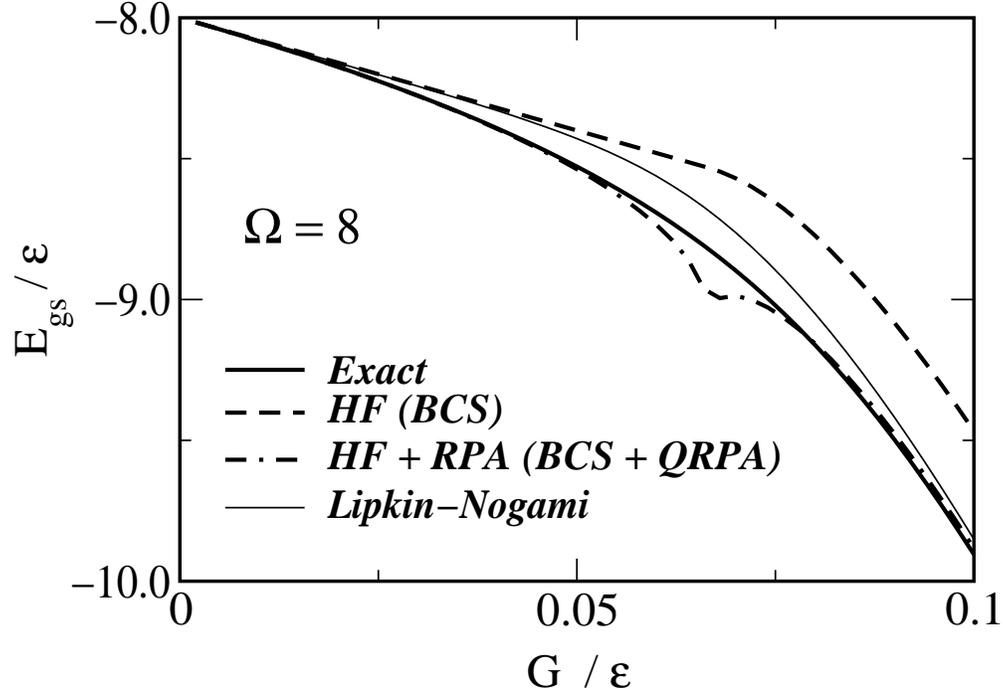}}
  \end{center}
\protect\caption{
Comparison of the ground state energy $E_{gs}$ obtained by several 
methods. 
The solid line is the exact numerical solution, while the ground state energy 
in the Hartree-Fock (BCS) approximation is denoted by the dashed line. 
The dot-dashed line takes the RPA correlation energy into account in 
addition to the HF (BCS) energy. The results of the Lipkin-Nogami method are 
denoted by the thin solid line. 
}
\end{figure}

\newpage

\begin{figure}
  \begin{center}
    \leavevmode
    \parbox{0.9\textwidth}
           {\psfig{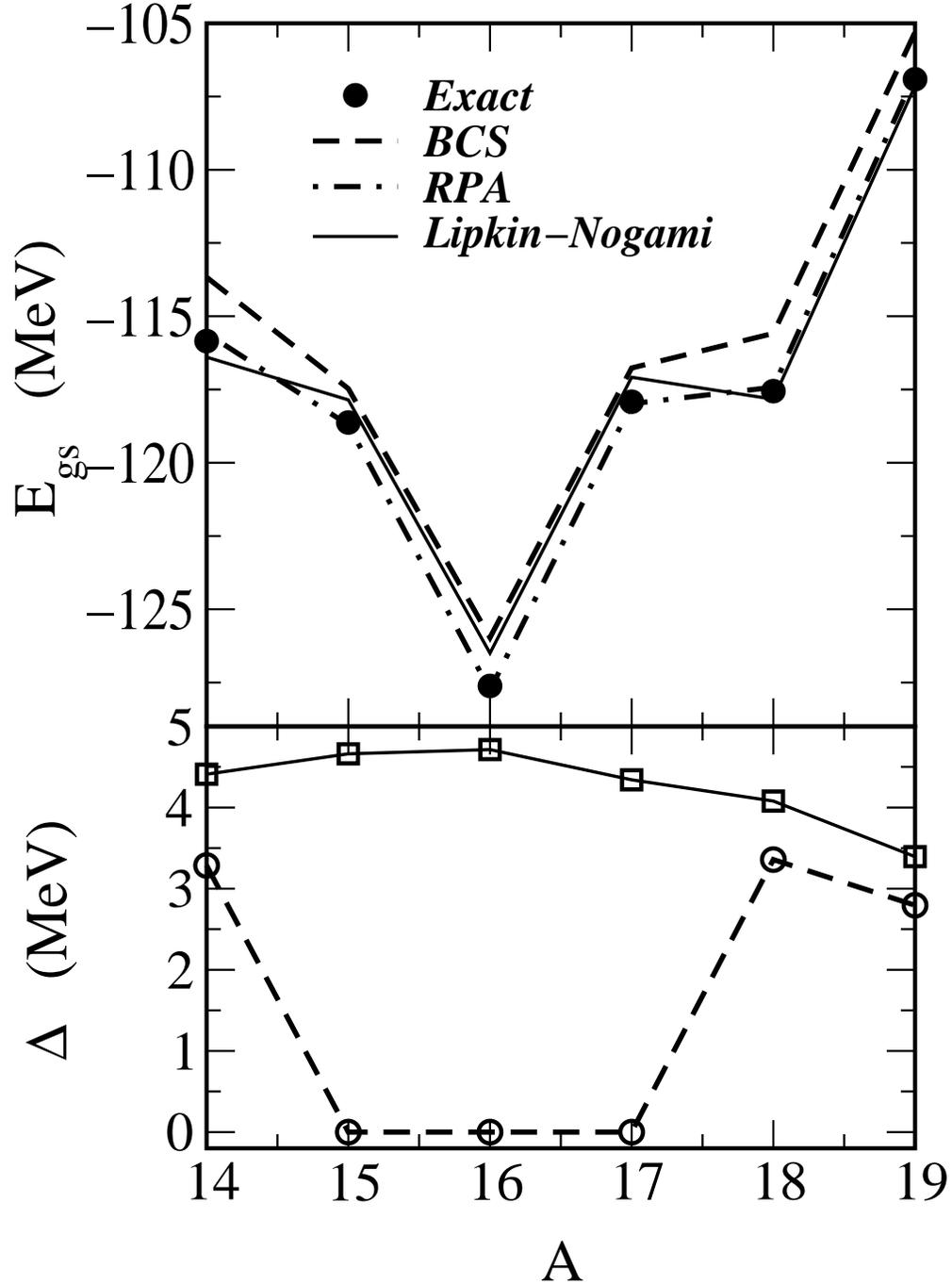}}
  \end{center}
\protect\caption{
The ground state energy $E_{gs}$ (the upper panel) and the 
pairing gap (the lower panel) for oxygen isotopes 
estimated with the two-level model 
as a function of the mass number.  
The exact reults are denoted by the filled circles, while 
the meaning of each line is the same as in Fig. 2. 
For the pairing gap in the Lipkin-Nogami method, $\lambda_2$ is added to 
the pairing gap $\Delta$. 
}
\end{figure}

\end{document}